\documentclass[preprint2]{aastex}
\usepackage{graphicx}
\usepackage{lineno}
\usepackage{lscape}
\usepackage{amsmath}
\usepackage[english]{babel}
\usepackage{times}
\usepackage{natbib}
\begin{document}

\title{High-Redshift Galaxies from Early JWST Observations: \\Constraints on Dark Energy Models } 

\author{N. Menci$^1$, M. Castellano$^1$, P. Santini$^1$, E. Merlin$^1$, A. Fontana$^1$, F. Shankar$^2$}
\affil{$^1$ INAF - Osservatorio Astronomico di Roma, via Frascati 33, I-00078 Monte Porzio, Italy}
\affil{$^2$School of Physics \& Astronomy, University of Southampton, Highfield, Southampton SO17 1BJ, UK}
\begin{abstract} 
Early observations with JWST have led to the discovery of an unexpected large density (stellar mass density $\rho_*\approx 10^{6}\,M_{\odot}\,Mpc^{-3}$) of massive galaxies (stellar masses $M_*\geq 10^{10.5}M_{\odot}$) at extremely high redshifts $z\approx 10$.  While such a result is based on early measurements which are still affected by uncertainties currently under consideration by several observational groups, its confirmation would have a strong impact on cosmology.  Here
we show that - under the most conservative assumptions, and independently of the baryon physics involved in galaxy formation - such galaxy abundance is not only in tension with the standard $\Lambda$CDM cosmology, but provides extremely tight constraints on the expansion history of the Universe and on the growth factors corresponding to a wide class of Dynamical Dark Energy (DDE) 
 models. Adopting a parametrization 
 $w=w_0+w_a(1-a)$ for the evolution of the DDE equation of state parameter $w$ with the expansion factor $a$, we derive constraints on combinations of $(w_0,  w_a)$
that  rule out with confidence level  $>2\sigma$ a major portion of the parameter space $(w_0,  w_a)$ allowed (or even favoured) by existing cosmological probes. 
\end{abstract}


\section{Introduction} 
The abundance of massive galaxies at high redshifts constitutes a powerful probe for cosmological models. In fact, in the standard Cold Dark Matter (CDM) scenario (see Peebles 1993) the exponential high-mass tail of the mass function of dark matter (DM) halos is expected to shift toward progressively smaller masses for increasing redshift (see, e.g., Del Popolo \& Yesilyurt 2007 for a review) at a rate that depends on the assumed cosmology. 
Hence, the comparison of the predicted abundance of massive DM haloes at increasingly larger redshift with the observed abundance of galaxies with corresponding stellar mass $M_*$  provides increasingly strong constraints on the assumed cosmological framework. 
 The critical issue in the comparison is the translation from a predicted number density $N(M,z)$ 
of DM halos with mass $M$ to a prediction for the abundance of galaxies with stellar mass $M_*$. In fact, the relation between  $M$ and  $M_*$ depends on the complex physics of baryons involved in galaxy formation. However, very robust constraints can be derived under the (extremely) conservative assumption  that the stellar mass corresponds to that of all the available baryons contained in a given DM halos $M_*=(\Omega_b/\Omega_m)\,M\equiv f_b\,M$,  where 
 $f_b$ is the cosmic baryon fraction. In fact, for any observed stellar mass $M_*$, assuming smaller $M_*/M$ ratios would yield larger  
 DM mass $M$, and hence lower predicted abundance. Thus, given an observed number density $N_{obs}(M_*,z)$ at redshift $z$, the condition $N(M_*/f_b,z)\geq N_{obs}(M_*,z)$  provides a robust and extremely conservative constraint on the halo mass function (and hence on cosmology) which is independent on the actual baryon physics relating $M_*$ and $M$. Such approach has been adopted in Menci et al. (2020) to constrain Dynamical Dark Energy (DDE) models  from different observations concerning the abundance of massive galaxies at redshifts $z=3-6$, and recently by Boylan-Kolchin (2022) and Lovell et al. (2022) to the abundance of massive galaxies measured from the 
  James Webb Space Telescope (JWST) NIRCam observations of the Cosmic Evolution Early Release Science (CEERS) program (Labb\'e et al. 2022). While calibration issues and uncertainties related to the  assumed 
  initial mass function (IMF) are still under debate, a confirmation of such 
 JWST measurements would imply that  the high mass end of the mass function evolves surprisingly little from $z\approx 10$ to $z\approx 6$, yielding a high stellar mass density $\rho_*\approx 10^{6}\,M_{\odot}\,Mpc^{-3}$ at $z=10$ which is in strong tension with the faster evolution predicted by $\Lambda$CDM. Here we apply the above approach to show that the observed  CEERS abundance is actually in tension with a wide class of cosmological models. Adopting for the DE equation of state parameter $w$ the form  $w=w_0+w_a(1-a)$ (introduced by Chevallier-Polarski-Linder, Chevallier and Polarski 2001, Linder 2003)  employed in most DE studies, we 
  derive strong constraints on the combinations $(w_0, w_a)$. These 
 have been shown to capture the dynamics of a wide class of scalar field DE models (for the mapping of such a parameterisation onto physical DE see, e.g, Caldwell and Linder 2005; Linder 2006; Scherrer 2015; Sangwan, Mukherjee,  Jassal 2017), although such a parametrisation fails to describe the  dynamics of some DE models, like those characterized by rapid, step-like transition of $w$ at $z\leq 2$ (see 
Linden \& Virey 2008).
  
\section{Method} 
\noindent We adopt the Sheth and Tormen (2001) mass function 
\begin{equation}
{dN\over dM}={A\,\overline{\rho}\over M^{2}}\,{d ln\, \nu\over d\,\,ln M}\,\Bigg({1 \over \overline{\nu}^{2q}}+1\Bigg)\,  {\overline{\nu}^{2}\over \pi}     e^{-\overline{\nu}^{2}/2} 
\end{equation}
where $\overline{\rho}$ is the background average density, $\nu=\delta_c/\sigma(M,z)$ where $\delta_c$ corresponds to the critical linear overdensity for collapse,  $\sigma(M ,z)$ is the variance of the linear density field smoothed on the scale 
$R=[3M/4\pi\overline{\rho}]^{1/3}$, and evolving with time according to the linear growth factor $D(z)$ of density perturbations. We assume a CDM form for the linear power spectrum (consistent with the measurements from the Cosmic Microwave Background, CMB, see Plank collaboration 2020). 
The parameters $a=0.71$ and $q=0.3$  are related to the physics of collapse, 
and $\overline{\nu}=\sqrt{a}\nu$. The normalization factor is $A=0.32$ 

The motivations for adopting the above form of the mass function are the following: 1) among the different proposed forms (see  Jenkins et al. 2001; Warren et al. 2006; Tinker et al. 2008) since the seminal paper by Press \& Schechter (1974), this is the expression which provides the most extended high-mass tail, and thus constitutes the most conservative form for our scopes; 2) theoretical works (Sheth et al. 2001, Maggiore and Riotto 2010, Corasaniti and Achitouv, 2011; Achitouv and Corasaniti 2012) have shown that its form is physically motivated in terms of the  collapse process of halos; 3) 
The Sheth and Tormen (2001) form has been tested  against N-body simulations for a variety of CDM cosmologies. These include the cases of a critical Universe, of an open Universe, and the $\Lambda$CDM case. Achitouv et al. (2014) studied the mass function in  Ratra-Peebles quintessence model of DE (Peebles \& Ratra 1988), finding that the parameters defining the key quantities determining the coefficients $q$ and $a$ change by less than 5 \% when passing from $\Lambda$CDM to the quintessence cosmology for the large  masses $M\geq 10^{11}\,M_{\odot}$ relevant to this paper. Despali et al. (2016) have tested the above  mass function against the SBARBINE set of  N-body simulations for a variety of combinations of $\Omega_m$ (ranging from 0.2 to 0.4)  and $\Omega_{\Lambda}$ (ranging from 0.6 to 0.8). These authors concluded that - with the proper definition of halo - the Sheth \& Tormen (2001) mass function is universal as a function of redshift and cosmology to within $20\%$, an uncertainty that we consider in our analysis.   
 As for the threshold $\delta_c$, we notice that in principle this depends weakly on cosmology. Here we shall adopt the conservative value $\delta_c=1.65$ for all DDE models. This constitutes a lower bound for the possible values taken in different DDE cosmologies (Mainini, Maccio, Bonometto, Klypin 2003; Pace, Waizmann, Bartelmann 2010) thus maximizing the predicted abundance of  DM halos. 

The exponential cutoff in eq. (1) is critically determined by the  cosmic expansion rate and by the growth factor $D(z)$, which depend on the equation of state of DE $w(a)=w_0+w_a(1-a)$. 
In the above parametrization, the standard $\Lambda$CDM cosmology corresponds to $w_0=-1$ and $w_a=0$. The cosmic expansion rate $H(z)$ and cosmic volume $V_{w_0,w_a}(z)$ depend on $w_0$ and $w_a$ as recalled in Menci et al. (2020). As for the growth factor $D(z)$ we use the parametrization to the solutions given in Linder (2005) and reported in Menci et al. (2020). This reproduces the behaviour of the growth factor to within 0.1\%-0.5\% accuracy for a wide variety of DDE cosmologies (Linder 2005, Linder and Cahn 2007)  and allows for a rapid scanning of the parameter space of DDE models. 
We  normalize the amplitude of linear power spectrum in terms of $\sigma_8$, the the local ($z=0$) variance of the density field smoothed over regions of 8 $h^{-1}$ Mpc.   Present cosmological constraints based on Planck data, baryonic acoustic oscillations and type-Ia supernovae yield  $\sigma_8\approx 0.8$   for the $\Lambda$CDM cosmology; such a value vary by $\approx 2$ \% when different combinations  ($w_0, w_a$)  are assumed  (Ade et al. 2016; Di Valentino 2017; Mehrabi et al. 2018). To adopt a {\it conservative} approach, so as to maximize the predicted abundance of large-mass DM halos, we  adopt the value $\sigma_8=0.83$. 

\section{Results}

Here we compare the maximal stellar mass densities $\rho_{max,w_0,w_a}(>M_*)$ allowed by different combinations $(w_0,w_a)$ with the values 
$\rho_{obs}(>M_*)$ measured by Labb\'e et al. (2022). We focus on the most massive bin considered by the above authors corresponding to $M_*\geq 
\overline{M_*}=10^{10.5}\,M_{\odot}$ in the redshift range $z_1=9\leq z \leq z_2=11$, corresponding to a cosmic volume $V(z_1,z_2)$. 
The corresponding predicted {\it maximal} (i.e. assuming  $M=\overline{M_*}/f_b$) stellar mass density 
\begin{equation}
\rho_{max}(>\overline{M*})\equiv
 \int_{z_1}^{z_2} \int_{\overline{M*}\over f_b}^{\infty}\,{dN\over dM}\,f_b\,M\,dM\,{dV\over dz}{dz\over V(z_1,z_2)}
\end{equation}
is shown in fig. 1 for DDE cosmologies with fixed $w_0=-1$ and four different values of $w_a$, 
and compared with the measurement by Labb\'e et al. (2022) from the early JWST observations. 

\vspace{0.1cm}
\hspace{-1cm}
\scalebox{0.32}[0.32]{\rotatebox{0}{\includegraphics{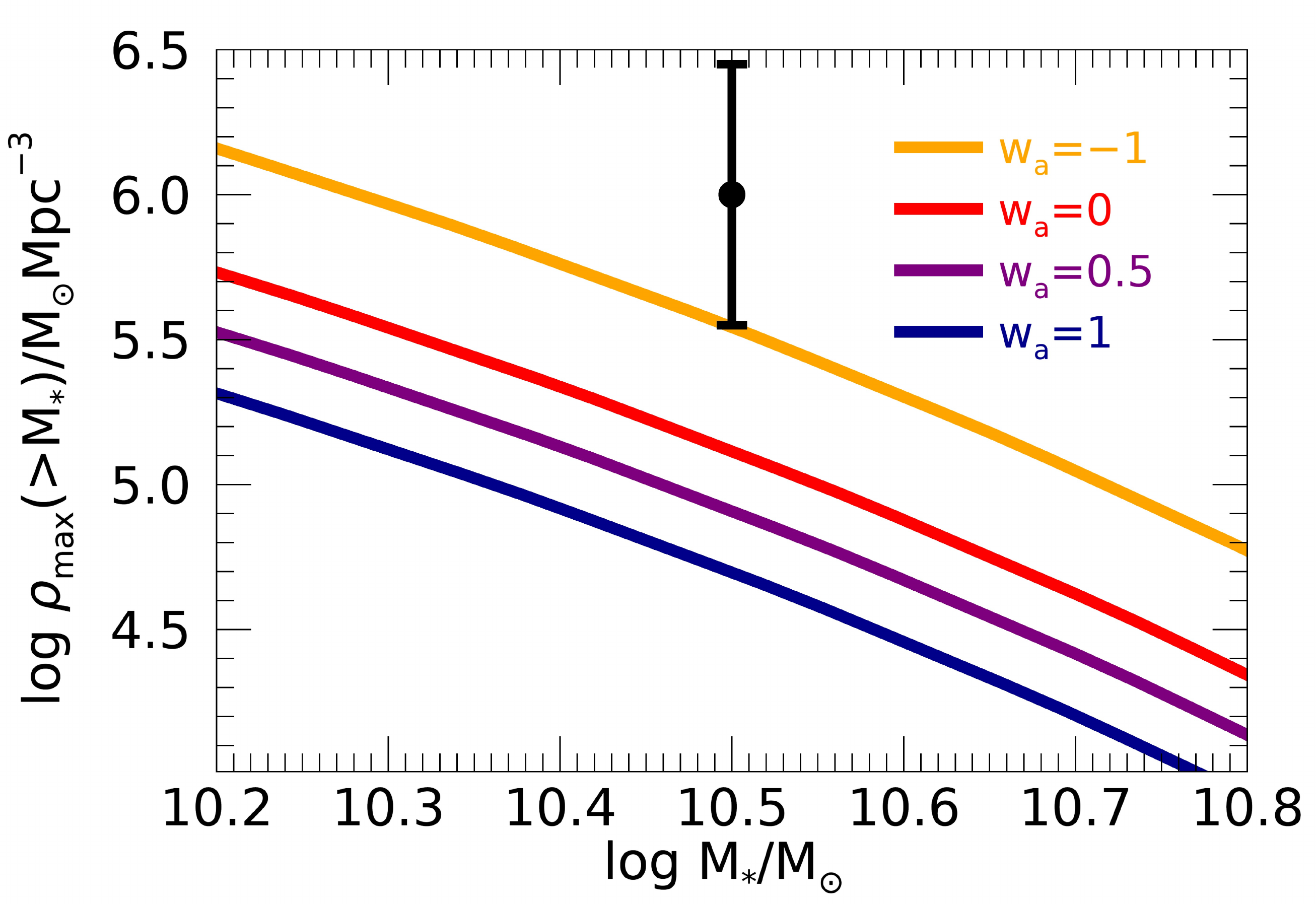}}}
\vspace{0.cm }
\newline
{\footnotesize Fig. 1. The maximal stellar mass density predicted by DDE scenarios with $w_0=-1$ and four different values of $w_a$ shown in the legend. In the predictions, we have considered an uncertainty of $0.5$ dex in the value of $M_*$. The point is the value measured by Labb\'e  et al. (2022). 
For the sake of simplicity, the latter - derived assuming a $\Lambda$CDM cosmology - has not been rescaled to the different values corresponding to different cosmological scenarios (see text). 
}
\vspace{0.cm }

To derive robust, conservative upper limits we assumed the most conservative value for $f_b$ which is still consistent with CMB observations. For the Planck values  (Planck Collaboration 2020) $\Omega_b\,h^{2}=0.0224\pm 0.0001$, $h\equiv H_0/100\,{\rm km\,s\,Mpc^{-1}}=0.674 \pm 0.05$, and $\Omega_m=0.315\pm 0.007$, we derive  $f_b=0.18$ as the most conservative estimate (this upper limit holds also for CMB measurements in DDE cosmologies, see Ade et al. 2016). 
In additions, in fig. 1 the predictions for $\log \rho_{max}$ from eq. 2 include the 20\% theoretical uncertainty on the mass function recalled in Sect. 2. This has been considered as an errorbar of 0.08 dex on the predictions for log $\rho_{max}$ computed from eq. 2, and the  predictions are conservatively computed at the upper tip of such errorbar.

The $\Lambda$CDM case
($w_a=0$) is well below the 1-$\sigma$ deviation from the observational value $\rho_{obs}$ and thus in tension (at $
\approx 2\sigma$ level) with observations as obtained by Boylan-Kolchin (2022). Notice that our prediction in the  $\Lambda$CDM case is slightly larger than that obtained by Boylan-Kolchin (2022) due to the our consideration of measurement  uncertainties in the observed value of $M_*$, and to our conservative choice of adopting $f_b=0.18$, while the latter author adopts $f_b=0.158$ derived from the best-fit values of the Plank cosmological parameters . 
The tension is larger for increasing values of $w_a$, thus showing that the condition $\rho_{max}(>M_*)\geq \rho_{obs}(>M_*)$ 
provides extremely stringent constraints on DDE models. 

To explore the impact of the measured stellar mass density on the full parameter space of DDE models, and derive proper 
confidence level for exclusion for each considered cosmology, 
we consider a  grid of DDE models characterized by different combinations $(w_0, w_a)$. For each combination we first correct  the observed densities $\rho_{obs}$ with the volume factor $f_{Vol}=V_{\Lambda}/V_{w_0,w_a}$ (computed in the redshift range $z=9-11$) to account for the fact that the volume density given in Labb\'e et al. (2022) have been derived  assuming a $\Lambda$CDM cosmology. Analogously, we must take into account  that the stellar masses measured by the above authors have been inferred from luminosities assuming a $\Lambda$CDM cosmology to convert observed fluxes into luminosities.  Thus, for each combination  $(w_0, w_a)$ we must correct the measured masses by a factor $f_{lum}=D^{2}_{L,w_0,w_1}/D^{2}_{L,\Lambda}$ where $D^{2}_{L,w_0,w_1}$ is the luminosity distance at $z=10$ 
 for the considered $(w_0,w_a)$ combination, and $D^{2}_{L,\Lambda}$ is its value in the $\Lambda$CDM  case. For each combination $(w_0, w_a)$ we compare the (cosmology corrected) observed mass density of galaxies $\rho_{obs}(>\overline{M_*})$ at $z=10$ with the predicted {\it maximal} density $\rho_{max,w_0,w_a}(>\overline{M_*})$ corresponding to each $(w_0, w_a)$ combination.  
 
\vspace{0.2cm }
\hspace{-1.5cm}
\scalebox{0.35}[0.35]{\rotatebox{0}{\includegraphics{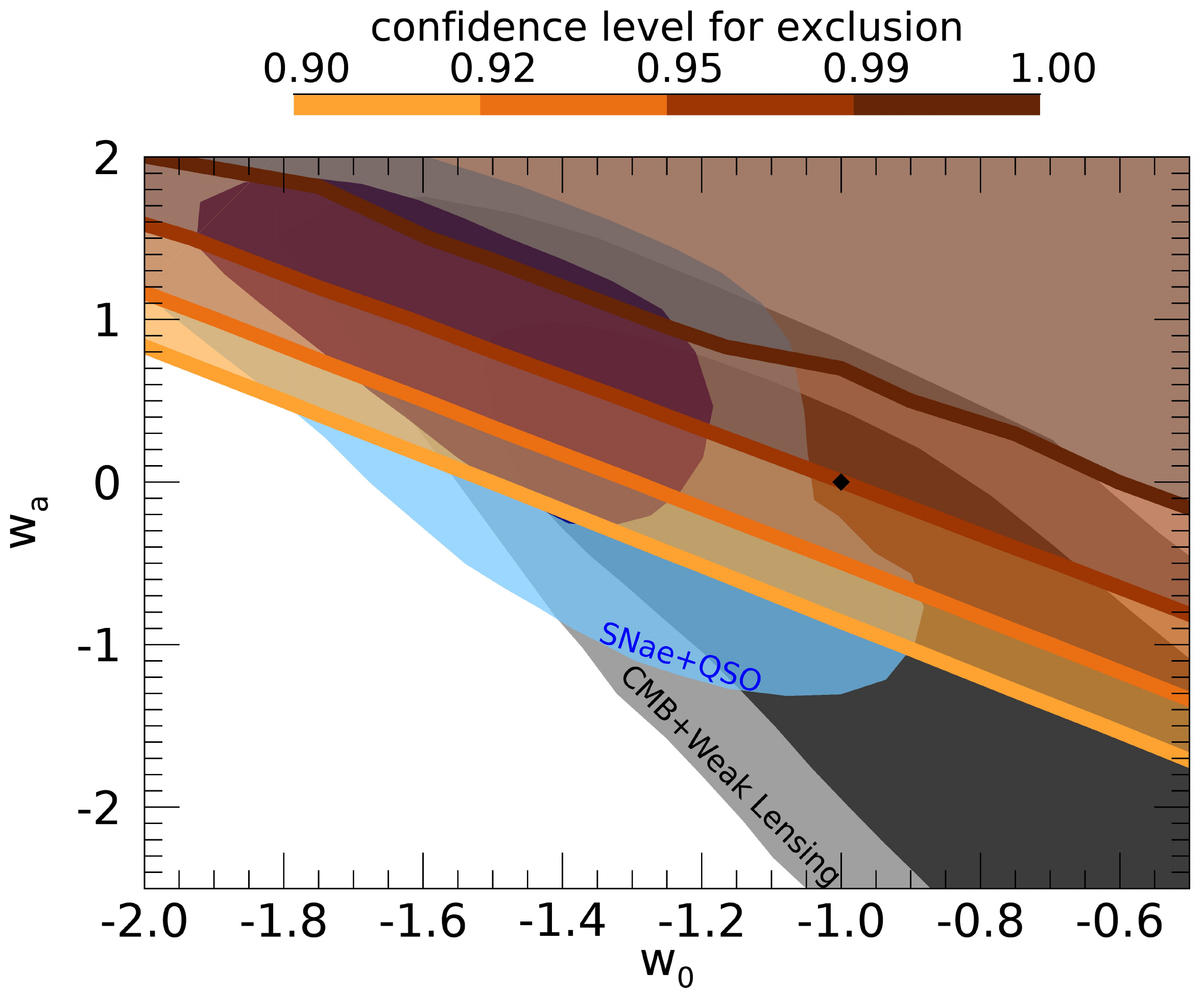}}}
\vspace{0.cm }
{\footnotesize Fig. 2.  Exclusion regions in the $w_0-w_a$ plane 
derived from the observed stellar mass density at $z=10$ (Labb\'e et al. 2022). 
The  regions above each colored line correspond to exclusion with the confidence levels shown in the upper bar. Our {\it exclusion} region is compared with the $2-\sigma$ and $3-\sigma$ contours {\it allowed} by  CMB+weak lensing (grey and dark-grey regions) and by  the combination of the same data with the Hubble diagram of supernovae and quasars (blue regions), derived from fig. 4 of Risaliti and Lusso (2019). The black dot corresponds to the $\Lambda$CDM case ($w_0=-1$, $w_a=0$).
}
\vspace{0.2cm } 
  
 The confidence for the exclusion  of each considered DDE model is obtained 
  as the probability  that  $\rho_{obs}(>\overline{M_*})>\rho_{max,w_0,w_a}(>\overline{M_*})$. The probability of 
 measuring a given value is derived through a Monte Carlo procedure based on the average value and variance given by the observed point and errorbars in Labb\'e et al. (2022). To be conservative, we assign an  errorbar of 0.5 dex to the measured stellar mass to account for systematic related to the SED fitting procedure (see Santini et a 2015). In our Monte Carlo procedure, the limit stellar mass $\overline{M_*}$ is extracted in such an interval after a flat distribution, to simulate systematic uncertainties. The resulting exclusion regions in the parameter space  $(w_0,w_a)$ is shown in fig. 2 for 
different confidence levels, and compared with regions allowed by existing probes. 

The $\Lambda$CDM case is excluded at almost 2-$\sigma$  level, while a major fraction of the parameter space is  excluded with high confidence level. The exclusion region is overplotted to the regions allowed by CMB and weak lensing constraints, and to the region derived 
by the combination of the same data with the Hubble diagram of supernovae and distant quasars (Risaliti and Lusso 2019). Our probe severely  restricts the  region in DDE parameter space allowed by other methods, and exclude almost all the region favored by the  distant quasar method. 
 
\section{Discussion and Conclusions}

Our results show the potential impressive impact of JWST observations of distant galaxies on cosmology. The measurements of the  abundance of galaxies
at very high redshifts within the reach of JWST indeed constitute a cosmological probe competitive with the existing canonical probes. 
We stress that, on the computational side, our results are extremely conservative and robust with respect to existing uncertainties, as we summarize below.
\newline 
$\bullet$ The results do not depend on the  physics of baryons involved in galaxy formation. Indeed, they have been derived under the extreme assumption that all baryon are in the form of stars. \newline 
$\bullet$  We adopt the Sheth \& Tormen halo mass function. Among the different forms proposed so far, this is the one that provides the more extended tail at high masses, and may even overestimate the abundance of massive halos (Wang et al. 2022)\newline
$\bullet$  Independence on the filter choice and on the form of DM. We adopt a CDM spectrum (Bardeen et al. 1986; Hlozek et al. 2012). However, on the large mass scales $M\geq 10^{10}\,M_{\odot}$ considered here the variance $\sigma(M)$ is in practice independent on the filter function used to relate it to the power spectrum (see Benson et al. 2013) and of the free-streaming properties of DM
\newline
$\bullet$  Conservative choice of cosmological parameters. For the normalization of the spectrum, we adopt $\sigma_8=0.83$, the upper limit allowed by Plank even considering alternative cosmologies (see Di Valentino 2017; Mehrabi et al. 2018), thus maximizing the predicted number of massive halos. For $h$, $\Omega_m$ and $\Omega_b$ we considered the combination that, within the uncertainties, allows for the largest  baryon fraction $f_b=0.18$. 
\newline 
$\bullet$ For the collapse threshold $\delta_c$ we chose the lower limit $\delta_c=1.65$ within the range allowed by DDE models, again maximizing the abundance of massive halos.
\newline
$\bullet$  We allowed for an additional uncertainty of 0.5 dex on the measured value $\overline{M_*}$. This again constitutes a conservative assumption. Uncertainties 
related to the SED fitting procedure reported in Labb\'e et al. (2022) correspond to values $\approx 0.2$ dex. 

While our conclusions are robust on the theoretical side, critical issues may affect the observational measurements we compare with. E.g., potential uncertainties may affect  the calibration of the JWST photometric data used by Labb\'e et al. (2022). Regrettably, there is not a firm assessment of the NIRCam calibrations, that may still be subject to revision to the level of 10-20\% especially in the short-wavelength bands ( Boyer et al. 2022). At face value, the aforementioned uncertainty is not expected to yield to revisions of the $M/L$ ratios of the targets large enough to affect significantly  our conclusions. A first test of the effects of the calibration revision using the independent  GLASS-JWST-ERS sample (Treu et al. 2022) showed  that on average stellar masses are affected to a level of $\approx 0.2$ dex. However, we caution that 
 the effect on the overall shape of the galaxy spectral energy distribution (SED) (as well as the assumptions  on the star formation histories adopted in the SED-fitting procedure, Ryan et al. 2022) may reflect in a non-linear way on the estimated physical parameters of some objects. 
A second issue may concern the Chabrier IMF  adopted by Labb\'e et al. (2022) to derive stellar masses. While assuming other universal forms for the IMF based on low-redshift conditions would not change (or even make stronger) the constraints we derive,  the star formation process can be significantly different at the  high redshifts. In particular, the increase of gas temperatures in star-forming, high-redshift galaxies  (also contributed by the heating due to CMB photons) could result in an increasing contribution of massive stars to the galactic light, which would yield significantly lower values for the stellar masses  (the exact value depending on the assumed gas temperature) compared to those measured by Labb\'e et al. (2022), see  Steinhardt et al. (2022). 

Our comprehensive statistical analysis reported in Fig. 2, highlights that only extreme DDE models, with very specific combinations of $w_0$ and $w_a$, may still be marginally consistent with existing cosmological probes. In addition, our results suggest that almost all baryons must be in the form of stars at high redshifts $z\approx 9-10$. The corresponding low gas fractions would lead to profound implications for Hydrogen reionization, owing to the expected large escape fractions, at least within the galactic stellar mass range probed in this work. 
  
\begin{acknowledgements}
\end{acknowledgements}

\end{document}